\documentclass[a4paper,twocolumn,prl]{revtex4-1}
\usepackage{amsmath}
\usepackage{amssymb}
\usepackage{graphicx}
\usepackage{amsthm}

\begin{document}

\newcommand{\ket}[1]{\left|#1\right>}
\newcommand{\bra}[1]{\left<#1\right|}
\newcommand{\sket}[1]{|#1\rangle}
\newcommand{\sbra}[1]{\langle#1|}

\newcommand{\braket}[2]{\langle#1|#2\rangle}
\newcommand{\ketbra}[2]{|#1\rangle\langle#2|}

\newcommand{\abs}[1]{\left|#1\right|}
\newcommand{\sabs}[1]{|#1|}

\newcommand{\norm}[1]{\left\|#1\right\|}
\newcommand{\snorm}[1]{\|#1\|}

\newcommand{\avr}[1]{\left<#1\right>}
\newcommand{\savr}[1]{\langle#1\rangle}

\title{Sufficient Condition for Entanglement Area Laws in Thermodynamically Gapped Spin Systems}
\author{Jaeyoon Cho}
\address{School of Computational Sciences, Korea Institute for Advanced Study, Seoul 130-722, Korea}
\date{\today}
\begin{abstract}
We consider general locally-interacting arbitrary-dimensional lattice spin systems that are gapped for any system size. We show under reasonable conditions that nondegenerate ground states of such systems obey the entanglement area law. In so doing, we offer an intuitive picture on how a spectral gap restricts the correlations that a ground state can accommodate and leads to such a special feature.
\end{abstract}
\maketitle

Over the last decade or so, quantum information theory has emerged as an indispensable tool in studying strongly-correlated many-body systems. For example, entanglement is essential in classifying quantum phases of matter~\cite{ost02,osb02,vid03}, especially topological quantum phases~\cite{kit06,lev06}, the density matrix renormalization group (DMRG) method provides the best way to numerically simulate low energy physics of one-dimensional spin systems~\cite{sch11}, and various universal features of general many-body systems have been explored with new tools and insights~\cite{nac06,has06c,bra06,eis06,vid08}.

While there are many different avenues in such quantum information approaches to many-body physics, they are essentially built upon the grounds of common theoretical foundations. The entanglement area law (or simply the area law) is one of the prominent~\cite{eis10}. For a many-body pure state $\sket{\Psi_0}$, the bipartite entanglement between a subregion ($A$) and the rest ($B$) is quantified by the entanglement entropy $S(\rho_A)$, the von Neumann entropy of the reduced density matrix $\rho_A=\text{Tr}_B\sket{\Psi_0}\sbra{\Psi_0}$. When $S(\rho_A)$ has an upper bound proportional to the surface area of $A$, we say $\sket{\Psi_0}$ obeys the area law~\cite{eis10}. It turns out that ground states of local Hamiltonians typically obey the area law~\cite{aud02,ple05,has07,wol08,bea10,got10,ara12,mic12,mic13,aco13,bra13}, possibly with a multiplicative logarithmic correction~\cite{vid03,wol06,mas09}, although one can deliberately construct a counterexample~\cite{vit10}. The area law is indeed a very special feature because in a large Hilbert space, almost all states, in the sense of the Haar measure, exhibit a volume-law scaling of the entanglement entropy; the states obeying the area law actually belong to a measure-zero set~\cite{pag93,pop06}. This anomaly leads to diverse and profound implications across various fields, e.g., in classical simulations of quantum systems~\cite{sch11}, topological quantum phases~\cite{kit06,lev06}, and Hamiltonian complexity theory~\cite{aha11}. Conceptually, the area law is also reminiscent of the holographic principle~\cite{bou02}. It has thus been of crucial importance to find out the general mechanism and criteria of the area law. In particular, one of the prominent open problems has been whether the area law is generally obeyed in gapped local systems in high dimension, since its one-dimensional problem was solved in Ref.~\cite{has07}. 

\begin{figure}
\includegraphics[width=0.9\columnwidth]{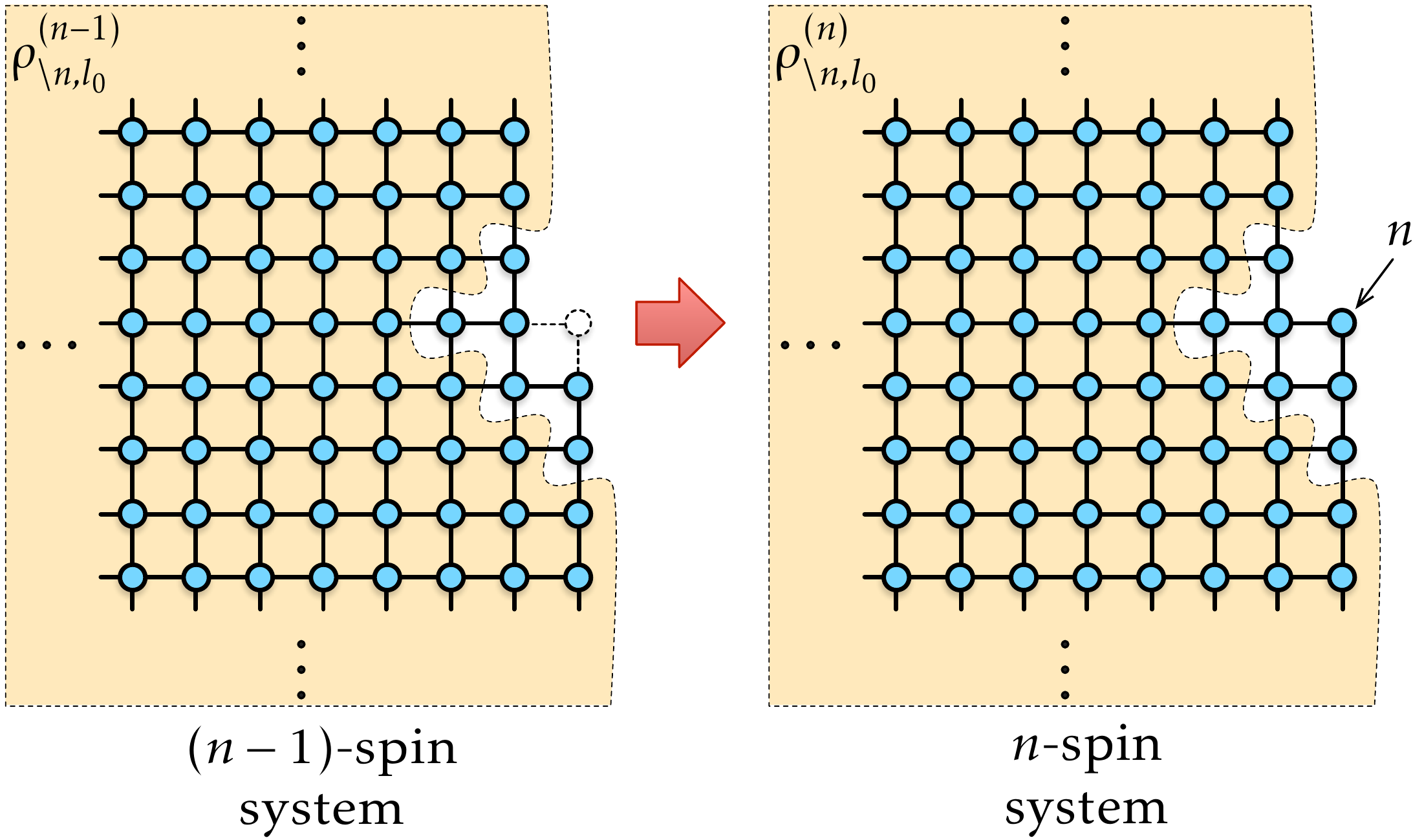}
\caption{The $(n-1)$-spin and the $n$-spin systems differ only locally. The reduced density matrices of the ground states $\rho_{\setminus n,l_{0}}^{(n-1)}$ and $\rho_{\setminus n,l_{0}}^{(n)}$ are obtained by tracing out a local neighborhood of spin $n$. In this illustration, $l_{0}=3$.}
\label{fig1}
\end{figure}

In this context, the fundamental question in hand is concerning the entanglement entropy of the ground state $\sket{\Psi_{0}^{(N)}}$ of a general local $N$-body Hamiltonian $H^{(N)}$ having a finite spectral gap $\Delta_{N}$ for sufficiently large $N$~\cite{has07}. As it turned out, in an arbitrary spatial dimension, proving (or disproving) the area law in such a general case is a daunting task at present. From a practical point of view, however, if we are in a position to tackle ordinary many-body systems (e.g., as in the context of classical simulations of quantum systems~\cite{sch11}),  we may bring in a few empirical assumptions without sacrificing much of the generality, thereby significantly relaxing the technical difficulties and furthermore offering a clear-cut insight into the problem. Specifically, we note that a many-body system is generally defined in terms of its microscopic details (i.e., its constituent particles, mutual interactions, external potential, etc.), while the system size $N$ is actually variable. Formally speaking, when $H^{(N)}$ is given, it is implicitly taken for granted that there also exist Hamiltonians $H^{(n)}$ with $n< N$ and importantly all different $H^{(n)}$ share the common defining characteristics of the system. For example, when we say a certain system is gapped, it generally means that $\Delta_{n}\ge\Delta$ for any $n$ with a lower bound $\Delta$. Here, all the different-sized ground states $\sket{\Psi_{0}^{(n)}}$ represent essentially the same matter and in the  thermodynamic limit, if exists, any local observable cannot discriminate between different $N$ as it becomes an intensive quantity. This makes it reasonable to assume that $\sket{\Psi_{0}^{(n)}}$ and $\sket{\Psi_{0}^{(n-1)}}$ have a finite overlap (i.e., they are nonorthogonal) after the boundary effect is properly washed out (see Fig.~\ref{fig1}) so that they share a common subspace that encapsulates the characteristic features of the system, e.g., the order parameters, correlation functions, and so on. Otherwise, the bulk properties of the system would be utterly unpredictable in practice as they will be drastically altered by a microscopic change of the system size, rendering the system unstable.

In this paper, we consider such a practical situation and prove the entanglement area law in arbitrary-dimensional gapped local spin systems under two general conditions drawn from the above reasoning. Before proceeding, let us first clarify our notation. We consider arbitrary systems of $N$ locally-interacting finite-dimensional spins placed on a $D$-dimensional lattice with one spin per site. For the `localness' of the interaction to make sense, the lattice has two properties. First, the Euclidean distance $\ell_E(s,s')$ and the graph distance $\ell_G(s,s')$ between sites $s$ and $s'$ satisfy $\ell_E(s,s')\le a_0\ell_G(s,s')$ for some constant $a_0$. Second, the number of sites in a unit volume $(\delta l)^D$ is bounded by $n_0(\delta l)^D$ for some constant $n_0$. For given site $s$, we define sets of neighboring sites as follows:
\begin{equation*}
\mathcal{B}_{s}^{k}=\{\text{site $s'$}:\ell_G(s,s')<k\}.
\end{equation*}
The interaction being local means that $N$-spin Hamiltonians can be written as a sum of local terms supported on $\mathcal{B}_s^{k_0}$ for $1\le s\le N$ with $k_0$ being a constant bounding the range of the interaction. The trace distance between density matrices $\rho$ and $\sigma$ is denoted by $\mathcal{D}(\rho,\sigma)=\frac{1}{2}\snorm{\rho-\sigma}_1\le1$ with $\snorm{\cdot}_1$ being the trace norm. The operator norm is denoted by $\snorm{\cdot}_\infty$.

The two conditions we impose are as follows. First, there exist a sequence of Hamiltonians with different systems sizes
\begin{equation*}
\{...,H^{(N-2)},H^{(N-1)},H^{(N)}\}
\end{equation*}
having nondegenerate ground states
\begin{equation*}
\{...,\sket{\Psi_0^{(N-2)}},\sket{\Psi_0^{(N-1)}},\sket{\Psi_0^{(N)}}\}
\end{equation*}
and finite spectral gap $\Delta_{n}\ge\Delta$ for all $n$. Here, we index the spins in such a way that the $n$-spin system is constructed by adding the $n$-th spin on the boundary of the $(n-1)$-spin system (see Fig.~\ref{fig1}). As the interaction is local, $H^{(n)}$ and $H^{(n-1)}$ differ by a local term
\begin{equation*}
K_n=H^{(n)}-H^{(n-1)}
\end{equation*}
supported on $\mathcal{B}_n^{2k_0}$ (not $\mathcal{B}_n^{k_0}$ in general because there can be distinct boundary terms). The interaction strength is finite, which means $\snorm{K_n}_\infty\le J$ for some constant $J$. 

Second, for all $n$, the ground states $\sket{\Psi_0^{(n)}}$ and $\sket{\Psi_0^{(n-1)}}$ have a finite overlap in the sense that there exist constants $\mu_0$ and $l_0$ such that
\begin{equation}
\mathcal{D}\left(\rho_{\setminus n,l_0}^{(n)},\rho_{\setminus n,l_0}^{(n-1)}\right)\le\mu_0<1,
\label{overlap}
\end{equation}
where $\rho_{\setminus s,k}^{(n)}=\text{Tr}_{\mathcal{B}_s^k}\sket{\Psi_0^{(n)}}\sbra{\Psi_0^{(n)}}$ (see Fig.~\ref{fig1}). For $l_{0}=1$, this condition means that when the ground state $\sket{\Psi_0^{(n)}}$ is written in terms of the eigenstates of $H^{(n-1)}$ and the states of spin $n$ as $\sket{\Psi_0^{(n)}}=\sum_{j}\alpha_j\sket{\Psi_j^{(n-1)}}\sket{\phi_j}_n$, the coefficient $\alpha_0$ associated with the ground state $\sket{\Psi_0^{(n-1)}}$ is non-zero. Although this seems intuitively natural, one can imagine a counterexample, albeit quite artificial, in which the entire system undergoes a quantum phase transition by a single change of the particle number at the boundary (see the Appendix). As mentioned, such an exceptional case is not of our interest in this work.

The main result of this paper is the following theorem. 

\newtheorem*{thm}{Theorem}
\begin{thm}
Consider a lattice spin system satisfying the above two conditions. Take a subregion $A$, which is a $D$-dimensional ball of radius $R_0$. For the ground state, the entanglement entropy of this region is bounded as
\begin{equation}
S(\rho_A)\le c_{D-1} R_0^{D-1}+c_{D-2} R_0^{D-2}+...+c_{1} R_0+c_0,
\label{arealaw}
\end{equation}
where $c_j$'s are constants determined by above-defined system parameters $\Delta$, $J$, $D$, $a_0$, $n_0$, $k_0$, $l_0$, and $\mu_0$. Hence, the upper bound of $S(\rho_A)$ scales as $R_0^{D-1}$, satisfying the area law.
\end{thm}

Here, we took the ball-shaped region to simplify the proof. Generalization to the case of a different shape is straightforward. Note that the area law makes sense only for simple-shaped regions. For example, the surface area of a fractal shape can be arbitrarily large.

\begin{figure}
\includegraphics[width=0.9\columnwidth]{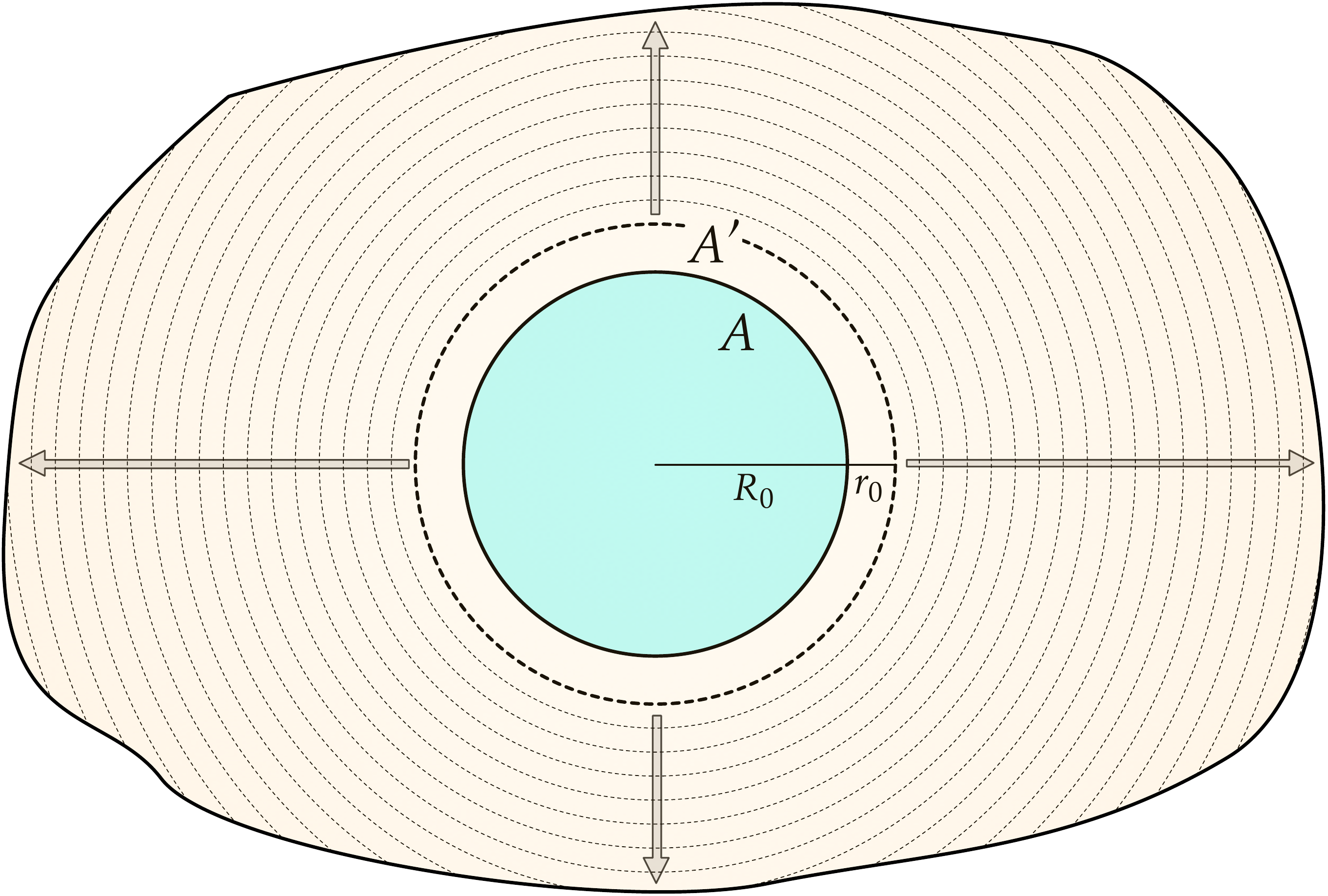}
\caption{Starting from a system corresponding to region $A'$, the system is gradually enlarged while retaining its shape as far as possible.}
\label{fig2}
\end{figure}

The underlying idea of the proof is as follows. Take a subregion $A'$ to be a $D$-dimensional ball of radius $R_0+r_0$ centered at the origin of region $A$ with positive constant $r_0\ll R_0$ to be chosen later (see Fig.~\ref{fig2}). Suppose there are $M$ spins in region $A$, $M+L$ spins in region $A'$, and $N-(M+L)$ spins in the rest. Our strategy is to take a particular sequence of ground states $\{\sket{\Psi_0^{(M+L)}},\sket{\Psi_0^{(M+L+1)}},...,\sket{\Psi_0^{(N)}}\}$ in the following way. (1) $\sket{\Psi_0^{(M+L)}}$ is the ground state of the $(M+L)$-spin system corresponding to region $A'$. (2) $\sket{\Psi_0^{(N)}}$ is the ground state of the whole system. (3) $\sket{\Psi_0^{(n)}}$ ($M+L<n<N$) is the ground state of  an intermediate system. Here we choose the $n$-th spin, among $N-(n-1)$ remaining ones, to be the one having the shortest Euclidean distance to the origin, whereby the shape of the system is retained as far as possible for all $n$ (see Fig.~\ref{fig2}). For each ground state in the sequence, we can obtain the reduced density matrix for region $A$,  $\{\rho_A^{(M+L)},\rho_A^{(M+L+1)},...,\rho_A^{(N)}\}$. Note that $S(\rho_A^{(M+L)})$ is bounded by the logarithm of the Hilbert space dimension for region $A'-A$. Our initial bound
\begin{equation}
\begin{split}
S(\rho_A^{(M+L)})\le L&\le n_0v_D[(R_0+r_0)^D-R_0^D]\\
&=n_0v_Dr_0DR_0^{D-1}+\mathcal{O}(R_0^{D-2})
\end{split}
\label{sinitial}
\end{equation}
thus exhibits the area-law scaling, where $v_D$ is the volume factor for $D$-dimensional balls. We thus find that
\begin{multline}
S(\rho_A)=S(\rho_A^{(N)})\\
\le S(\rho_A^{(M+L)})+\sum_{n=M+L+1}^N[S(\rho_A^{(n)})-S(\rho_A^{(n-1)})]
\label{sfinal}
\end{multline}
exhibits the area-law scaling if the summation on the right-hand side also does so for appropriately chosen $r_0$. In order to show this, we need to understand how $\sket{\Psi_0^{(n-1)}}$ is mapped to $\sket{\Psi_0^{(n)}}$. 

The ensuing procedure is based on the following intuitive picture. Suppose we try to identify $\sket{\Psi_0^{(n)}}$ with $U_n^{k}\sket{\Psi_0^{(n-1)}}\sket{\phi_0}_n$, where $U_n^k$ is a unitary operator supported on $\mathcal{B}_n^k$ and $\sket{\phi_0}_n$ is any state of spin $n$. 
In general, the exact identity would be obtained only for sufficiently large $k$, i.e., $\sket{\Psi_0^{(n)}}=U_n^{\infty}\sket{\Psi_0^{(n-1)}}\sket{\phi_0}_n$. However, one can see that the spectral gap can play a role here in approximating $U_{n}^{\infty}$. Note that both $H^{(n-1)}$ and $H^{(n)}$ are gapped. {\em If} $H(\lambda)=(1-\lambda)(H^{(n-1)}+S_n)+\lambda H^{(n)}=H^{(n-1)}+(1-\lambda)S_{n}+\lambda K_{n}$ also remains gapped for $0\le \lambda\le 1$, where $S_n=\Delta(I-\sket{\phi}_n\!\sbra{\phi})$, we can then consider an adiabatic passage from $\sket{\Psi_0^{(n-1)}}\sket{\phi_0}_n$ to $\sket{\Psi_0^{(n)}}$. Intuitively, $H(\lambda)$ is likely to be gapped for appropriately chosen $\sket{\phi_0}_n$ since only a small portion of the Hamiltonian is varied and $\sket{\Psi_0^{(n-1)}}$ and $\sket{\Psi_0^{(n)}}$ are essentially the same kind of states and hence no quantum phase transition occurs. For the moment, suppose it is the case. If so, as the adiabatic passage can be done in a finite time scale (inversely-proportional to $\Delta$) and the Hamiltonian is varied only locally, the Lieb-Robinson bound implies that the adiabatic process can affect the system only (quasi-)locally, which means that $U_n^\infty$ is approximately local. One can thus write $\sket{\Psi_0^{(n)}}=U_n^{x_0}\sket{\Psi_0^{(n-1)}}\sket{\phi_0}_n$ for some constant $x_0$ up to a small error. {\em If} we neglect the error for the moment, we can choose $r_0$ to satisfy $r_0\ge a_0x_0$ so that every term in the summation of Eq.~\eqref{sfinal} vanishes because $U_n^{x_0}$ does not act on region $A$, resulting in the area law. Conceptually, what happens is that the boundary effect spreads no further than $r_{0}$ away and thus $\rho_A^{(n)}$ converges to $\rho_A$ once $n$ reaches the point where region $A$ does not recognize the existence of a boundary any more. 

In the above picture, we have made two logical jumps to be resolved. First, it should be ensured that $H(\lambda)$ is indeed gapped. Second, $U_n^\infty$ is only approximately local and thus we need to work out how the errors add up. As a preliminary step, note that condition~\eqref{overlap} implies there is a unitary operator $V_{n}$ acting on $\mathcal{B}_n^{l_0}$ such that $\sabs{\sbra{\Psi_0^{(n)}}V_{n}\sket{\Psi_0^{(n-1)}}\sket{\phi_0}_n}\ge1-\mu_0>0$, which follows from the Uhlmann's theorem~\cite{nie00}. Let $H_1^{{(n)}}=V_{n}(H^{(n-1)}+S_n)V_{n}^\dagger$. This Hamiltonian preserves the gap condition and the ground state is $\sket{\xi_0^{(n)}}=V_{n}\sket{\Psi_0^{(n-1)}}\sket{\phi_0}_n$. For convenience, let $H_2^{(n)}=H^{(n)}$ and $\sket{\eta_0^{(n)}}=\sket{\Psi_0^{(n)}}$. Letting $H_0^{(n)}$ be the sum of all terms commonly appearing both in $H_1^{(n)}$ and $H_2^{(n)}$, we can write $H_{\{1,2\}}^{(n)}=H_0^{(n)}+h_{\{1,2\}}^{(n)}$, where $h_{\{1,2\}}^{(n)}$ are supported on $\mathcal{B}_n^{l_0+k_0}$.

We are now ready to proceed. We have local Hamiltonians $H_{\{1,2\}}^{(n)}=H_0^{(n)}+h_{\{1,2\}}^{(n)}$, which have a gap lower-bounded by $\Delta$. Their ground states are $\sket{\xi_0^{(n)}}$ and $\sket{\eta_0^{(n)}}$, respectively, and $\sabs{\braket{\xi_0^{(n)}}{\eta_0^{(n)}}}\ge1-\mu_0>0$. If we can find an adiabatic path from $\sket{\xi_0^{(n)}}$ to $\sket{\eta_0^{(n)}}$, we can also find one from $\sket{\Psi_0^{(n-1)}}\sket{\phi_0}_n$ to $\sket{\Psi_0^{(n)}}$ up to a local unitary operator $V_{n}$ that does not affect $S(\rho_A^{(n)})$ as long as $r_0\ge a_0l_0$. The key lemma for our proof is the following.

\newtheorem*{lm}{Lemma}
\begin{lm}
Introduce an ancillary two-level system $a$ and consider a local Hamiltonian
\begin{equation*}
\tilde{H}^{(n)}(\lambda)=H_s^{(n)}(\lambda)+h_a^{(n)}(\lambda),
\end{equation*}
where
\begin{equation*}
\begin{split}
H_s^{(n)}(\lambda)=&H_0^{(n)}+[h_1^{(n)}+\lambda\Delta]\otimes\sket{1}_a\sbra{1}\\
&\quad\quad\,+[h_2^{(n)}+(1-\lambda)\Delta]\otimes\sket{2}_a\sbra{2},\\
h_a^{(n)}(\lambda)=&f(\lambda)\Delta(\sket{1}_a\sbra{2}+\sket{2}_a\sbra{1})
\end{split}
\end{equation*}
with real non-negative smooth function $f(\lambda)$ for $0\le\lambda\le1$. $f(\lambda)$ vanishes at $\lambda=0$ and $\lambda=1$, and is maximized at $\lambda=1/2$ with $f(1/2)=f_0>0$. There exists $f(\lambda)$ with $f_0=\frac{1}{10}(1-\mu_0)$ such that $\tilde{H}^{(n)}(\lambda)$ is gapped for all $\lambda$ and the minimal gap is at least $\tilde\Delta=f_0(1-\mu_0)\Delta$. 
\end{lm}

\begin{figure}
\includegraphics[width=0.9\columnwidth]{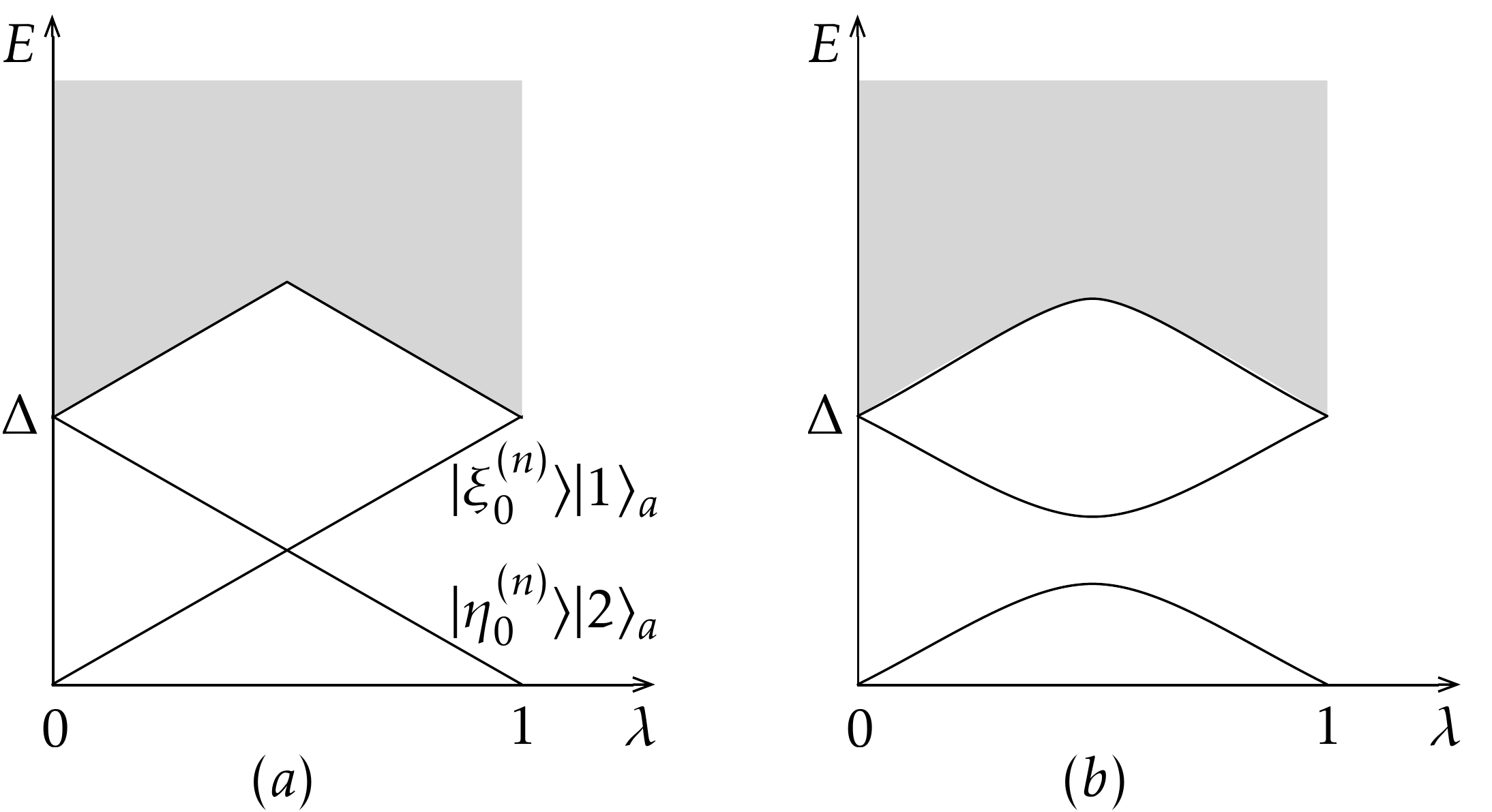}
\caption{(a) Energy spectrum of $H_{s}^{(n)}(\lambda)$.  (b) By adding a coupling term, the ground state degeneracy is lifted.}
\label{fig3}
\end{figure}

The underlying idea for the proof of this lemma is as follows. Let $h_{a}^{(n)}(\lambda)=0$ for the moment. $H_{s}^{(n)}(\lambda)$ alone is then readily diagonalized by $\sket{\xi_{j}^{(n)}}\sket{1}_{a}$ and $\sket{\eta_{j}^{(n)}}\sket{2}_{a}$, where $\sket{\xi_{j}^{(n)}}$ and $\sket{\eta_{j}^{(n)}}$ are the eigenstates of $H_{1}^{(n)}$ and $H_{2}^{(n)}$, respectively. As shown in Fig.~\ref{fig3}(a), the two lowest energy levels $\sket{\xi_{0}^{(n)}}\sket{1}_{a}$ and $\sket{\eta_{0}^{(n)}}\sket{2}_{a}$ become degenerate only at $\lambda=1/2$. This degeneracy can be lifted by adding a term that couples the two levels. $h_{a}^{(n)}(\lambda)$ plays this role as $\sabs{\braket{\xi_0^{(n)}}{\eta_0^{(n)}}}>0$. The detailed proof is presented in the Appendix. Note that the ground state of $\tilde{H}^{(n)}(0)$ is $\sket{\xi_0^{(n)}}\sket{1}_a$ and that of $\tilde{H}^{(n)}(1)$ is $\sket{\eta_0^{(n)}}\sket{2}_a$. The lemma thus implies that there exists an adiabatic path from $\sket{\Psi_0^{(n-1)}}\sket{\phi_0}_n\sket{1}_a$ to $\sket{\Psi_0^{(n)}}\sket{2}_a$ up to an irrelevant local unitary transformation, where the Hamiltonian varies locally on $\mathcal{B}_n^{l_0+k_0}$ and $a$. 

Such a local adiabatic change preserves the area law, as discussed in Ref.~\cite{aco13}. We tailor their method in the remainder of our proof. Let $\sket{\tilde{\Psi}_0^{(n)}(\lambda)}$ be the ground state of $\tilde{H}^{(n)}(\lambda)$. By slightly modifying the derivation in Ref.~\cite{osb07}, one can construct an artificial Hamiltonian governing the change of $\sket{\tilde{\Psi}_0^{(n)}(\lambda)}$ in $\lambda$, which turns out to be approximately local in the following sense.

\newtheorem*{adia}{Exact adiabatic evolution}
\begin{adia}
There exists an integer constant $\tilde{l}>l_0+k_0$ such that
\begin{equation}
i\frac{d}{d\lambda}\sket{\tilde{\Psi}_0^{(n)}(\lambda)}=\Bigl[F_{\tilde{l}}^{(n)}(\lambda)+\sum_{j\ge\tilde{l}+1}G_j^{(n)}(\lambda)\Bigr]\sket{\tilde{\Psi}_0^{(n)}(\lambda)},
\label{adiabatic}
\end{equation}
where $F_{\tilde{l}}^{(n)}(\lambda)$ and $G_j^{(n)}(\lambda)$ are Hermitian, $F_{\tilde{l}}^{(n)}(\lambda)$ is supported on $\mathcal{B}_n^{\tilde{l}}$, $G_j^{(n)}(\lambda)$ is supported on $\mathcal{B}_n^j$, and furthermore $\snorm{G_j^{(n)}(\lambda)}_\infty<g_0(j-l_{0}-k_{0})^{-4D}$ for some constant $g_0$. Here, $\tilde{l}$ and $g_{0}$ are determined by $\tilde{\Delta}$, $\max_{\lambda}\sabs{df(\lambda)/d\lambda}\Delta$, and $\snorm{h_{\{1,2\}}}_{\infty}\le Jn_{0}v_{D}a_{0}^{D}(l_{0}+k_{0})^{D}$.
\end{adia}

The last ingredient of our proof is the small incremental entangling theorem presented in Ref.~\cite{aco13}.

\newtheorem*{sie}{Small incremental entangling theorem}
\begin{sie}
Consider a many-body system in a pure state $\ket\psi$. The system is divided into four regions $A_1$, $A_2$, $A_3$, $A_4$, and evolves by a Hamiltonian $H_{23}$ supported on region $A_2+A_3$. The reduced density matrix for region $A_1+A_2$ at time $t$ is given by $\rho_{12}(t)=\text{Tr}_{34}e^{-iH_{23}t}\sket{\psi}\sbra{\psi}e^{iH_{23}t}$. The growth rate of the entanglement entropy $S(\rho_{12}(t))$ at any $t$ is bounded as
\begin{equation*}
\frac{d}{dt}S(\rho_{12}(t))\le c_e\snorm{H_{23}}_\infty \log[\min(d_2,d_3)],
\end{equation*}
for some constant $c_e>0$, where $d_{j}$ is the Hilbert space dimension for region $A_{j}$.
\end{sie}

We are now ready to finish up our proof. Let us choose $r_{0}>a_{0}\tilde{l}$. We can bound $S(\rho_{A}^{(n)})-S(\rho_{A}^{(n-1)})$ as follows. Suppose spin $n$ has Euclidean distance $R_{0}+r$ to the origin of region $A$ with $r\ge r_{0}$. As $\sket{\tilde{\Psi}_0^{(n)}(0)}$ and $\sket{\tilde{\Psi}_0^{(n)}(1)}$ yield entanglement entropies $S(\rho_{A}^{(n-1)})$ and $S(\rho_{A}^{(n)})$ for region $A$, respectively, we can examine $\sket{\tilde{\Psi}_0^{(n)}(\lambda)}$ to obtain the bound. Note that $\sket{\tilde{\Psi}_0^{(n)}(1)}$  is obtained by evolving $\sket{\tilde{\Psi}_0^{(n)}(0)}$ through Eq.~\eqref{adiabatic} during unit time.  By performing the Trotter expansion, one realizes that only $G_j^{(n)}(\lambda)$'s with $a_0j\ge r$ can affect the entanglement entropy. By using the small incremental entangling theorem, we find that 
\begin{equation*}
\begin{split}
S&(\rho_{A}^{(n)})-S(\rho_{A}^{(n-1)})\\
&\le c_{e}\sum_{j\ge r/a_{0}}\snorm{G_{j}^{(n)}(\lambda)}_{\infty}\times(\text{number of spins in $\mathcal{B}_{n}^{j}$})\\
&\le c_eg_0\sum_{j\ge r/a_0}\frac{n_{0}v_{D}a_{0}^{D}j^{D}}{(j-l_0-k_0)^{4D}}\\
&\le c_{e}g_{0}n_{0}v_{D}a_{0}^{D}\int_{r/a_{0}}^{\infty}\frac{(x-1)^{D}}{(x-1-l_{0}-k_{0})^{4D}}dx.
\end{split}
\end{equation*}
It thus follows that the summation on the right-hand side of Eq.~\eqref{sfinal} is bounded by
\begin{multline*}
\int_{r_{0}}^{\infty}dr\int_{r/a_{0}}^{\infty}dx\, n_{0}v_{D}D(R_{0}+r)^{D-1}\\
\times c_{e}g_{0}n_{0}v_{D}a_{0}^{D}\frac{(x-1)^{D}}{(x-1-l_{0}-k_{0})^{4D}},
\end{multline*}
which is $\mathcal{O}(R_{0}^{D-1})$ (note that $r_{0}/a_{0}>1+l_{0}+k_{0}$). $c_{j}$ in Eq.~\eqref{arealaw} can be obtained by expanding this integral and Eq.~\eqref{sinitial}, which completes the proof of the theorem.

As a final remark, we note that our approach in its current form  is not applicable to topologically ordered systems in general as our inherent physical setting based on an open boundary condition and nondegenerate ground states is not compatible with the nontrivial topology of the space and the topological degeneracy, which are the essential attributes of topological quantum phases~\cite{kit06,lev06}. For such systems governed by frustration-free Hamiltonians,  the local topological quantum order, if exists, leads to the area law~\cite{mic13}. The general proof without such restrictions is however yet to be given.

\section*{Appendix}

\subsection{I. Contrived gapped system violating condition (1)}

Consider a system of 4-dimensional spins with basis states $\{\ket1,\ket2,\ket3,\ket4\}$. Suppose the Hamiltonian is written as
\begin{equation*}
\begin{split}
H^{(n)}=&H_{a}^{{(n)}}+H_{b}^{(n)}\\
&+\Delta\sum_{s>1}(\ket1_{s-1}\!\bra1+\ket2_{s-1}\!\bra2)(\ket3_{s}\!\bra3+\ket4_{s}\!\bra4)\\
&+\Delta\sum_{s>1}(\ket3_{s-1}\!\bra3+\ket4_{s-1}\!\bra4)(\ket1_{s}\!\bra1+\ket2_{s}\!\bra2),
\end{split}
\end{equation*}
where $H_{a}^{(n)}$ contains only $\{\ket1,\ket2\}$, while $H_{b}^{(n)}$ contains only $\{\ket3,\ket4\}$. Both $H_{a}^{(n)}$ and $H_{b}^{(n)}$ are gapped with a minimal gap $\Delta$, but the ground state energy of $H_{a}^{(n)}$ is zero for $n<M$ and $\Delta$ for $n\ge M$, whereas that of $H_{b}^{(n)}$ is $\Delta$ for $n<M$ and zero for $n\ge M$. The ground state is then spanned by $\{\ket1,\ket2\}$ for $n<M$ and by $\{\ket3,\ket4\}$ for $n\ge M$. In this case, the overlap between $\sket{\Psi_{0}^{(n<M)}}$ and $\sket{\Psi_{0}^{(n\ge M)}}$ vanishes for any $l_{0}$.

\subsection{II. Remarks on the boundary effect and the thermodynamic limit}

Consider a mapping from $\sket{\Psi_{0}^{(n-1)}}$ to $\sket{\Psi_{0}^{(n)}}$, as in Fig.~1 of the main text. As the Hamiltonian $H^{(n)}$ differs from $H^{(n-1)}$ only by a local term $K_{n}$, it is reasonable to approximate $\sket{\Psi_{0}^{(n)}}$ with $U_n^{k}\sket{\Psi_0^{(n-1)}}\sket{\phi_0}_n$ by finding an optimal unitary operator $U_n^{k}$ supported on $\mathcal{B}_n^k$ so that the approximation gets better as $k$ is increased. 
Let us define a function
\begin{equation*}
\mu_{n}(k)=\inf_{U_{n}^{k}}\frac{1}{\sqrt{2}}\norm{\sket{\Psi_{0}^{(n)}}-U_n^{k}\sket{\Psi_0^{(n-1)}}\sket{\phi_0}_n}\le1.
\end{equation*}
Apparently, $\mu_{n}(k)$ is a non-increasing function of $k$ and $\lim_{k\rightarrow\infty}\mu_{n}(k)=0$. We can also define
\begin{equation*}
\mu(k)=\sup_{n}\mu_{n}(k),
\end{equation*}
which is also non-increasing. 
One can think of $\mu(k)$ as a characteristic function of the system that quantifies how far the boundary effect penetrates into the bulk. This is also intimately related to the existence of a thermodynamic limit. 
For instance, if $\mu(k)=0$ for $k>k_{B}$ with $k_{B}$ being a certain constant, $\rho_{\setminus n,k}^{(n-1)}$ becomes identical to $\rho_{\setminus n,k}^{(n)}$ for $k>k_{B}$ (see Fig. 1 of the main text). Consequently, if one takes any region sufficiently away from the boundary, the corresponding reduced density matrix becomes independent of the system size and hence all the local quantities of the bulk become intensive quantities, i.e., there exists a thermodynamic limit of the system at the zero temperature. Ordinarily, $\mu(k)$ would be a certain {\em decreasing} function, whose functional form characterizes how fast the system converges to the thermodynamic limit. 
Note that the condition~(1) of the main text is violated if and only if $\mu(k)=1$ for all $k$ (the system in the previous section is an example).
For gapped systems satisfying the condition~(1), i.e., $\mu(l_{0})<1$ for some $l_{0}$, 
one can show that $\mu(k)$ asymptotically decreases at least exponentially, and hence the thermodynamic limit should exist (this will be discussed elsewhere).

\subsection{III. Proof of the lemma}

\begin{figure}
\includegraphics[width=\columnwidth]{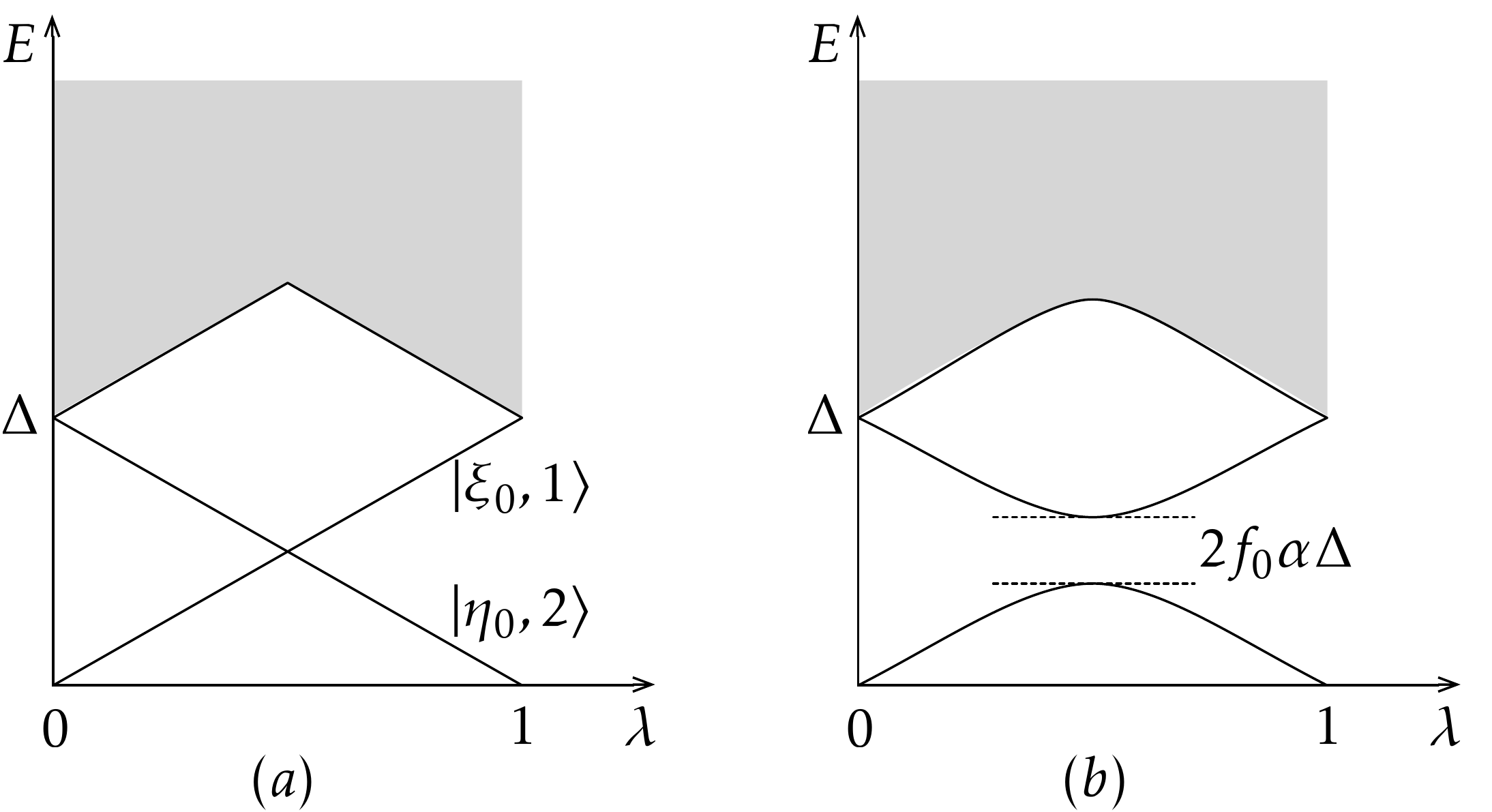}
\caption{Energy spectrums of (a) $H_{s}(\lambda)$ and (b) $H_{g}(\lambda)$.}
\label{fig4}
\end{figure}

Throughout the proof, we omit superscript $(n)$ for brevity. Let
\begin{equation*}
(H_0+h_1)\sket{\xi_j}=p_j\sket{\xi_j}
\end{equation*}
with eigenvalues $p_j$ and eigenstates $\sket{\xi_j}$. Without loss of generality, $p_0=0$ and $p_j\ge\Delta$ for $j\ge1$. In the same manner, let 
\begin{equation*}
(H_0+h_2)\sket{\eta_j}=q_j\sket{\eta_j}
\end{equation*}
with $q_0=0$ and $q_j\ge\Delta$ for $j\ge1$. $H_s(\lambda)$ is readily diagonalized with eigenstates $\sket{\xi_j}\sket{1}_a$ and $\sket{\eta_j}\sket{2}_a$ with eigenvalues $p_j+\lambda\Delta$ and $q_j+(1-\lambda)\Delta$, respectively. For brevity, let us write the state as $\sket{\xi_j,1}=\sket{\xi_j}\sket{1}_a$ and similarly for others. Fig.~\ref{fig4}(a) depicts the spectrum of $H_s(\lambda)$. The two lowest energy levels $\sket{\xi_0,1}$ and $\sket{\eta_0,2}$ cross only at $\lambda=1/2$. Consequently, by adding an additional term that couples $\sket{\xi_0,1}$ and $\sket{\eta_0,2}$, one can open a gap at $\lambda=1/2$, making the Hamiltonian gapped for all $\lambda$. $h_a(\lambda)$ can do this as $\sabs{\sbra{\xi_0,1}h_a(\lambda)\sket{\eta_0,2}}\ge f_0(1-\mu_0)\Delta>0$ for $\lambda=1/2$. If $1-\mu_0$ is close to one, it would be easy to find $f(\lambda)$ making $\tilde{H}(\lambda)$ gapped. If $1-\mu_0$ is very small, however, it is not necessarily obvious because $\sket{\xi_0,1}$ and $\sket{\eta_0,2}$ mostly couple to higher energy levels, whereas their mutual coupling is very small. 

Without loss of generality, choose the phase so that $\alpha=\braket{\xi_0}{\eta_0}\ge1-\mu_0$ is real. Let
\begin{equation*}
P_0=\sket{\xi_0,1}\sbra{\xi_0,1}+\sket{\eta_0,2}\sbra{\eta_0,2}.
\end{equation*}
A part of the Hamiltonian
\begin{equation*}
\begin{split}
P_0 \tilde{H}(\lambda)P_0=&\lambda\Delta\sket{\xi_0,1}\sbra{\xi_0,1}+(1-\lambda)\Delta\sket{\eta_0,2}\sbra{\eta_0,2}\\
&+f(\lambda)\alpha\Delta(\sket{\xi_0,1}\sbra{\eta_0,2}+\sket{\eta_0,2}\sbra{\xi_0,1})
\end{split}
\end{equation*}
has eigenvalues
\begin{equation*}
\omega_\pm(\lambda')=\Delta\left(\frac{1}{2}\pm\sqrt{\lambda'^2+f(\lambda')^2\alpha^2}\right)
\end{equation*}
with $\lambda'=\lambda-1/2$. Denote by $\sket{\omega_\pm(\lambda')}$ the corresponding eigenstate. Let 
\begin{equation*}
H_\perp(\lambda')=(1-P_0)\tilde{H}(\lambda')(1-P_0).
\end{equation*}
It is easy to see that for any state $\sket{\varphi}$ with $(1-P_0)\sket{\varphi}=\sket{\varphi}$, $\sabs{\sbra{\varphi}H_\perp(\lambda')\sket{\varphi}}\ge (3/2-\sabs{\lambda'}-f(\lambda'))\Delta$. Hereafter, we omit $\lambda'$ for brevity when the meaning is clear from the context (e.g., instead of $f(\lambda')$, we simply write $f$). We can write the Hamiltonian as
\begin{equation*}
\begin{split}
\tilde{H}=&\omega_-\sket{\omega_-}\sbra{\omega_-}+\omega_+\sket{\omega_+}\sbra{\omega_+}+H_\perp\\
&+(1-P_0)h_aP_0+P_0 h_a(1-P_0).
\end{split}
\end{equation*}
Let $H_g$ and $H_f$ denote, respectively, the first and the second line of this Hamiltonian. $H_g$ has a gap $2\Delta\sqrt{\lambda'^2+f^2\alpha^2}$ (see Fig.~\ref{fig4}(b)). Note that $\snorm{H_f}_\infty=f\Delta$. Let $\tilde{H}\sket{\Psi_j}=E_j\sket{\Psi_j}$ with eigenvalues $E_j$ and eigenstates $\sket{\Psi_j}$ for $j\ge0$. In what follows, we prove that
\begin{equation*}
E_1-E_0\ge f_0\alpha\Delta
\end{equation*} 
for all $\lambda'$ if 
\begin{equation*}
\begin{cases}
f(\lambda')=f_{0} & \text{for }\sabs{\lambda'}\le f_{0}\alpha,\\
f_{0}^{2}\le f(\lambda')\le f_{0} & \text{for }f_{0}\alpha<\sabs{\lambda'}\le\frac{1}{5},\\
0\le f(\lambda')\le f_{0}^{2} & \text{for }\sabs{\lambda'}>\frac{1}{5}.\\
\end{cases}
\end{equation*}

\subsubsection{1. For $\sabs{\lambda'}\le1/5$}

We regard $H_f$ as a perturbation to $H_g$. As $\sbra{\omega_\pm}H_f\sket{\omega_\pm}=0$, the first-order perturbation to $\omega_\pm$ vanishes. Moreover, as $\sbra{\omega_+}H_f\sket{\omega_-}=0$, the second-order perturbation to $\omega_\pm$ is at most $\sim f_0^2\Delta$, which can be much smaller than the unperturbed gap $\sim2f_0\alpha\Delta$ if $f_0$ is sufficiently small. 

In order to obtain a rigorous bound, let us write the perturbed eigenstate of $\sket{\omega_\pm}$ as $\sket{\Psi}=x\sket{\omega_-}+y\sket{\omega_+}+z\sket{\omega_\perp}$, where $\braket{\omega_-}{\omega_\perp}=\braket{\omega_+}{\omega_\perp}=0$, $\sabs{x}^2+\sabs{y}^2+\sabs{z}^2=1$, and the perturbed eigenvalue $E=\Delta/2+\epsilon$. We then have
\begin{multline*}
H(x\sket{\omega_-}+y\sket{\omega_+}+z\sket{\omega_\perp})\\
=(\Delta/2+\epsilon)(x\sket{\omega_-}+y\sket{\omega_+}+z\sket{\omega_\perp}),
\end{multline*}
which leads to
\begin{equation*}
\begin{split}
x\omega_{-}+z\sbra{\omega_-}h_a\sket{\omega_\perp}&=x(\Delta/2+\epsilon),\\
y\omega_{+}+z\sbra{\omega_+}h_a\sket{\omega_\perp}&=y(\Delta/2+\epsilon),\\
z\sbra{\omega_\perp}H_\perp\sket{\omega_\perp}+x\sbra{\omega_\perp}h_a\sket{\omega_-}+y\sbra{\omega_\perp}&h_a\sket{\omega_+}\\
&=z(\Delta/2+\epsilon).
\end{split}
\end{equation*}
From this set of equations, we find
\begin{multline*}
\sbra{\omega_\perp}H_\perp\sket{\omega_\perp}-(\Delta/2+\epsilon)\\
=\frac{\sabs{\sbra{\omega_+}h_a\sket{\omega_\perp}}^2}{\Delta\sqrt{\lambda'^2+f^2\alpha^2}-\epsilon}-\frac{\sabs{\sbra{\omega_-}h_a\sket{\omega_\perp}}^2}{\Delta\sqrt{\lambda'^2+f^2\alpha^2}+\epsilon}.
\end{multline*}
Note that the left hand side is larger than $(1-\abs{\lambda'}-f)\Delta-\epsilon\ge 7\Delta/10-\epsilon$ and that $\sqrt{\lambda'^{2}+f^{2}\alpha^{2}}\ge f_{0}\alpha$.
Note also that when $\epsilon=0$, the left hand side is larger than the right hand side.
Consequently, the lowest-energy solution $E_{0}$ exists when
\begin{equation*}
\Delta\sqrt{\lambda'^2+f^2\alpha^2}+\epsilon<0
\end{equation*}
and the second lowest solution $E_{1}$ is positive. We thus find 
\begin{equation*}
E_1-E_0>\Delta\sqrt{\lambda'^2+f^2\alpha^2}\ge f_{0}\alpha\Delta.
\end{equation*}

\subsubsection{2. For $\sabs{\lambda'}>1/5$}

Note that
\begin{equation*}
E_0\le\sbra{\omega_-}\tilde{H}\sket{\omega_-}=\omega_-.
\end{equation*}
Let us write $\sket{\Psi_0}=\sqrt{1-c_0^2}\sket{\omega_-}+c_0\sket{\omega_\perp}$ with real $c_0\le1$, where $\braket{\omega_-}{\omega_\perp}=0$. It follows that
\begin{equation*}
\begin{split}
E_0=\sbra{\Psi_0}\tilde{H}\sket{\Psi_0}=&(1-c_0^2)\omega_-+c_0^2\sbra{\omega_\perp}H_g\sket{\omega_\perp}\\
&+2c_0\sqrt{1-c_0^2}\text{Re}[\sbra{\omega_-}H_f\sket{\omega_\perp}]\\
\ge&(1-c_0^2)\omega_-+c_0^2\omega_+-2c_0\sqrt{1-c_0^2}f\Delta.
\end{split}
\end{equation*}
These two inequalities lead to
\begin{equation*}
c_0^2\le\frac{4f^2\Delta^{2}}{(\omega_+-\omega_-)^2+4f^2\Delta^{2}}.
\end{equation*}
Similarly, write $\sket{\Psi_1}=c_1\sket{\omega_-}+\sqrt{1-c_1^2}\sket{\omega_\perp'}$ with real $c_1<1$ and $\braket{\omega_-}{\omega_\perp'}=0$. For $\braket{\Psi_0}{\Psi_1}$ to vanish, we require
\begin{equation*}
c_1\sqrt{1-c_0^2}=c_0\sqrt{1-c_1^2}\sabs{\braket{\omega_\perp}{\omega_\perp'}}\le c_0\sqrt{1-c_1^2},
\end{equation*}
which results in $c_1\le c_0$. It thus follows that
\begin{equation*}
\begin{split}
E_1=&\sbra{\Psi_1}\tilde{H}\sket{\Psi_1}\\
\ge&c_1^2\omega_-+(1-c_1^2)\omega_+-2c_1\sqrt{1-c_1^2}f\Delta\\
\ge&\omega_+-\frac{4f^2\Delta^{2}}{(\omega_+-\omega_-)^2+4f^2\Delta^{2}}(\omega_+-\omega_-)-f\Delta.
\end{split}
\end{equation*}
Consequently, we end up with
\begin{equation*}
E_1-E_0\ge\frac{(\omega_+-\omega_-)^2}{(\omega_+-\omega_-)^2+4f^2\Delta^{2}}(\omega_+-\omega_-)-f\Delta.
\end{equation*}
It is easy to see that this bound is larger than $f_0\alpha\Delta$ for $\sabs{\lambda'}>1/5$.


%

\end{document}